\documentclass[pre,preprint,showpacs,epsf,epsfig]{revtex4}
\usepackage{graphicx}
\usepackage{graphics}
\usepackage{wrapfig}
\usepackage{dcolumn}
\usepackage{soul}
\usepackage{amssymb}
\usepackage{bm}
\usepackage{epsfig}
\usepackage{psfrag}
\usepackage{color}
\usepackage[utf8x]{inputenc}
\usepackage{ucs}
\usepackage{amsmath}
\usepackage{amsfonts}
\usepackage{amssymb}
\usepackage{graphicx}% Include Figure files
\usepackage{dcolumn}% Align table columns on decimal point
\usepackage{bm}% bold math
\usepackage{latexsym}
\usepackage{hyperref}

\def\3dots{\:\raisebox{-0.5ex}{$\stackrel{\textstyle.}{:}$}\:}
\def\beq{\begin{equation}}
\def\eeq{\end{equation}}
\def\bea{\begin{eqnarray}}
\def\eea{\end{eqnarray}}

\DeclareMathOperator{\tr}{Tr}

\begin{document}
\title{Catapulting of topological defects through elasticity bands in active nematics}

\author{Nitin Kumar$^{1,2,3,4,\ddag}$}
\author{Rui Zhang$^{3,5,\ddag}$}
\author{Steven A. Redford$^{1,6, \ddag}$}
\author{Juan J. de Pablo$^{7,8}$}
\email{depablo@uchicago.edu}
\author{Margaret L. Gardel$^{1,2,7}$}
\email{gardel@uchicago.edu}
\affiliation{
	$^1$James Franck Institute, The University of Chicago, Chicago, Illinois 60637, USA \\
    $^2$Department of Physics, The University of Chicago, Chicago, Illinois 60637, USA\\
    $^3$Institute for Biophysical Dynamics, The University of Chicago, Chicago, Illinois 60637, USA\\
    $^4$Department of Physics, Indian Institute of Technology Bombay, Mumbai, India\\
    $^5$Department of Physics, Hong Kong University of Science and Technology, Clear Water Bay, Kowloon, Hong Kong SAR\\
    $^6$Graduate Program in Biophysical Sciences, The University of Chicago, Chicago, Illinois 60637, USA\\
    $^7$Pritzker School of Molecular Engineering, The University of Chicago, Chicago, Illinois 60637, USA \\
    $^8$Institute for Molecular Engineering, Argonne National Laboratory, Lemont, Illinois 60439, USA\\
    $\ddag$ \textit{These authors contributed equally to this work}
    }
\date{\today}
%\draft
\begin{abstract}
 Active materials are those in which individual, uncoordinated local stresses drive the material out of equilibrium on a global scale. Examples of such assemblies can be seen across scales from schools of fish to the cellular cytoskeleton and underpin many important biological processes. Synthetic experiments that recapitulate the essential features of such active systems have been the object of study for decades as their simple rules allow us to elucidate the physical underpinnings of collective motion. One system of particular interest has been active nematic liquid crystals (LCs). Because of their well understood passive physics, LCs provide a rich platform to interrogate the effects of active stress. The flows and steady state structures that emerge in an active LCs have been understood to result from a competition between nematic elasticity and the local activity. However most investigations of such phenomena consider only the magnitude of the elastic resistance and not its peculiarities. Here we investigate a nematic liquid crystal and selectively change the ratio of the material's splay and bend elasticities. We show that increases in the nematic's bend elasticity specifically drives the material into an exotic steady state where elongated regions of acute bend distortion or "elasticity bands" dominate the structure and dynamics. We show that these bands strongly influence defect dynamics, including the rapid motion or "catapulting" along the disintegration of one of these bands thus converting bend distortion into defect transport.   Thus, we report a novel dynamical state resultant from the competition between nematic elasticity and active stress.
 
\end{abstract}
\maketitle

\section{INTRODUCTION}

Active materials are those in which components locally break detailed balance \cite{SRReview,RMP}. This local energy injection coupled with both local and global dissipative mechanisms---which depend on the exact material properties---leads to complex dynamical states \cite{ SRReview,needleman_dogic_nrm2017,Hatwalne,ViscReduc,ViscInc,MesoTurbulence,NKFlock,HowFar,PUndefined,CatesMIPS,TrapNK,battle2016broken}. Understanding how the interplay between local force generation and specific material viscoelasticity control the emergent structure and dynamics is an outstanding challenge in active matter.  Generalized hydrodynamic approaches have proven successful in understanding active fluids at large length and time scales \cite{RMP,JoannyProst}.  However, the peculiarities of specific systems can lead to exotic dynamical states at more immediate scales.  

Due to their intrinsic and tunable elasticity, nematic liquid crystals (LC) are an ideal system to answer questions about the interplay of elastic dissipation and active driving \cite{deGennes_book}. Formed from dense packings of rod-like constituents, LCs elastically resist distortions to their mesoscopic order while remaining locally fluid like.  When an extensile active stress outcompetes this nematic restoring force, the material is deformed.  Sufficient deformation of this kind results in the nucleation of $\pm$ 1/2 topological defect pairs \cite{Simha,Giomi,ThampiEPL,decamp2015orientational}. Once defects are created, LCs driven by extensile (contractile) stresses are bend (splay) unstable which results in the propulsion of comet-like +1/2 defects along (against) their orientation. At sufficiently high activity level, an ``active turbulent'' state develops. In this regime, topological defects constantly form, move, and annihilate creating complex hydrodynamic flows \cite{vjScience,Dogic,NKumar,thampi2014vorticity}. In well developed turbulence, the dynamical properties of the active nematic are well described by hydrodynamic models \cite{giomi2013,thampi2013,zhang2016,NKumar,thampi2014active}. These models show that when elastic force is much less than the active force, hydrodynamic effects dominate the dynamics. At lower activity levels, however, one expects elasticity to play an greater role in the dynamics and for emergent structural patterns to potentially deviate from those predicted in the hydrodynamic limit. Indeed, one example of such a structure is found in transient, elongated bend deformations which have been observed at low activity levels in numerical and experimental systems generally preceding from globally aligned initial states \cite{Aronson,Sagues,Ali,patteli,chandrakar2020confinement,senoussi2019tunable,nejad2021memory}. 

Here we explore structure and dynamics in active nematics where active and elastic stresses are similar in magnitude.  We exploit a highly tunable biomolecular nematic LC that is comprised of the biopolymer F-actin and driven by the molecular motor myosin II \cite{NKumar}. In this system, modulation of nematic elasticity can be achieved either through control over F-actin length, $l$, or by the addition of small quantities of microtubule biopolymers, which are 100-fold stiffer than F-actin  \cite{RZhang}. Further, time-dependent variation in myosin II concentration facilitates exploration of nematic structure and dynamics over a large range of activities. Recently, we showed that, for high activity, this system is well described by a hydrodynamic model of active nematics \cite{NKumar}. However, for LC with high bend elasticity and at intermediate activity, we observe a  dynamical steady state comprised of elongated "bands" of bend deformations. These “elasticity bands” undergo continuous creation and destruction, with highly variable length and width. The presence and persistence of these structures is highly dependent on nematogen length, indicating an important role of nematic elasticity in their formation. To isolate the contributions of splay and bend elasticity, we perform experiments where we selectively modify bend elasticity and find its increase to be sufficient to control band formation. Measuring the elastic distortion within the nematic as a function activity, we find that bands observed at intermediate activity coincide with a state of maximal elastic distortion. This likely reflects the increased energetic barrier for defect creation in LC with high bend elasticity. This observation is confirmed with hydrodynamic simulations, which also reveal that the elastic distortion of high elasticity LC increases non-monotonically with activity in a history dependent manner.  Finally, we show that the elastic distortion stored in bands is relieved by a rapid motion, or ``catapulting", of +1/2 defects. Taken together, our results demonstrate a new dynamic steady state in active nematics governed by the interplay of activity and bend elasticity.

\section{Results}
%\subsection{Observation of elasticity bands in F-actin based active nematics}
\begin{figure*}[h!]
	\centerline{\includegraphics[width=0.8\textwidth]{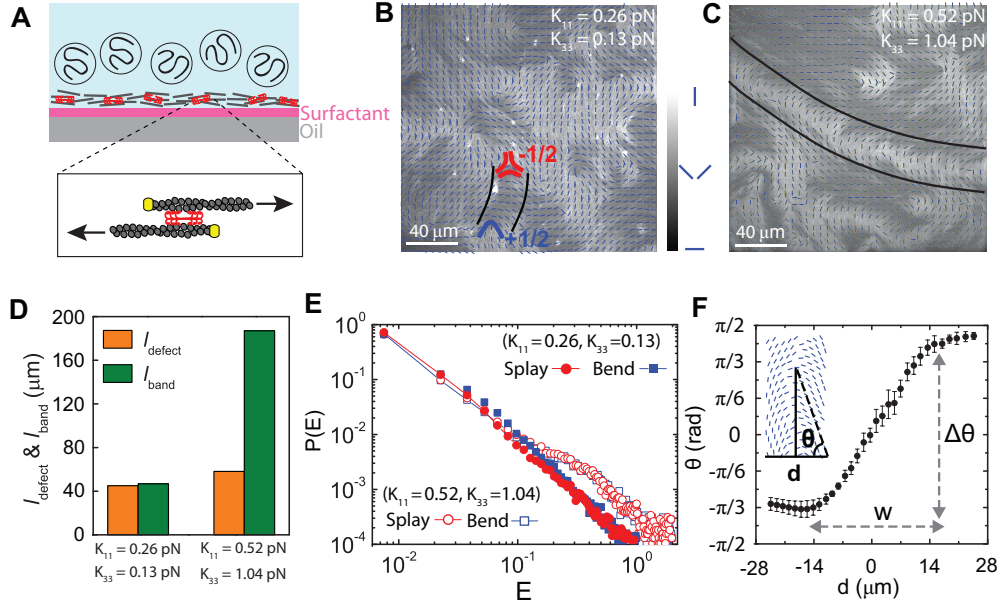}}
	\caption{\textbf{Observation and Characterization of elasticity bands in active nematics}: (\textbf{A}) Schematic representation of the experiment showing F-actin (grey) crowded to an oil-water interface using methylcellulose (circles). Inset shows myosin II motors (red) translocating short F-actin, with capping protein (yellow) indicating the F-actin barbed end. (\textbf{B}) Images of fluorescent F-actin in LC of low elasticity ($K_{11}$ = 0.26 pN, $K_{33}$ = 0.13 pN) (B) and high elasticity LC ($K_{11}$ = 0.52 pN, $K_{33}$ = 1.04 pN) (C). The local director field superposed with the colorbar showing how the intensity variations map to the local nematic field. A pair of $\pm$1/2 defects are separated by a region of bend deformation (band) outlined by solid black lines. (\textbf{D}) Maximum band length ($l_{band}$) and mean defect spacing ($l_{defect}$) for the two nematics shown above. (\textbf{E}) Probability distribution of the bend and splay elastic distortion for the two nematics described above. The higher elasticity nematic, plotted with open symbols, exhibits a heavy tail corresponding to bands. (\textbf{F}) The variation of the director field across an elastic band as shown in the inset. The band width, $w$, is determined by distance over which the director field orientation $\theta$ changes linearly.}
	\label{fig1}
\end{figure*} 

A thin nematic LC was formed by crowding short ($\sim$ $\mu$m) F-actin to a surfactant-stabilized oil-water interface using the depletant methylcellulose and waiting 30-45 minutes (Fig. 1A).  The average F-actin length ($l$) is controlled by varying the concentration of F-actin capping protein (CP) \cite{CP,KimPNAS}.  Our previous work showed that the splay ($K_{11}$) and bend ($K_{33}$) elasticity of such a nematic can be varied as $\sim l$ and $\sim l^3$, respectively \cite{RZhang}. Utilizing this tool, we investigate the activity of liquid crystals with three different elasticities: $l$ = 1 $\mu$m ($K_{11}$ = 0.26 pN, $K_{33}$ = 0.13 pN),  $l$ = 1.5 $\mu$m ($K_{11}$ = 0.44 pN, $K_{33}$ = 0.56 pN) and $l$ = 2 $\mu$m ($K_{11}$ = 0.52 pN, $K_{33}$ =1.04 pN) \cite{RZhang}. Active stress is introduced to the system by the addition of the molecular motor myosin-II. Upon their addition, myosin-II motors slide antiparallel actin filaments past each other resulting in net extensile stresses in the LC. This activity results in the spontaneous creation of topological defect pairs and emergent complex flows with a typical active state shown in Fig. \ref{fig1}B (Movie S1). The optical contrast in these panels arises from the polarized excitation laser stimulating fluorophores that label the length of the actin filament \cite{kinosita,sase}. Thus, this optical contrast provides a direct readout of the local nematic director field up to a symmetry factor \cite{zhang2021spatiotemporal}. The bright and dark regions of the image correspond to actin filaments aligned in the vertical and horizontal directions, respectively (see colorbar in Fig. \ref{fig1}B). From these images, the local nematic director field is determined \cite{NKumar}, and indicated by the blue lines. The local nematic director field is utilized to identify important features such as $\pm$ 1/2 topological defects, points of vanishing nematic order, in the LC. Such a defect pair is shown in Fig. \ref{fig1}B with a schematic of their shapes. 

In addition to point-like topological defects we notice--in certain cases--extended two dimensional structures that we term "elasticity bands". Elasticity bands, or simply bands, are extended regions between a pair of defects that are perpendicular to the surrounding director field. An example of an elasticity band in a nematic with $K_{11}$ = 0.26 pN and $K_{33}$ = 0.13 pN is outlined by black solid lines in Fig. \ref{fig1}B. To understand the structure of these interesting features we measure the director field orientation across a line segment locally perpendicular to the band with coordinate $d$. A plot of the director field angle  $\theta$ as a function of $d$ is shown in Fig. \ref{fig1}F and reveals a region where $\theta$ varies linearly as a function of $d$ flanked by regions of constant $\theta$. We then can define the width as the distance over which theta varies linearly. One can also measure the length of a band as the distance between the flanking defects. When we do so for nematics of varying elasticity we notice that higher elasticity LCs sport bands of dramatically increased length as seen in Fig. \ref{fig1}C (Movie S2). The band length, $l_{band}$, and defect spacing, $l_{defect}$, for nematics at low and high elasticity are shown in Fig. \ref{fig1}D. For nematics with shorter filaments and thus lower elasticity, the band length is similar to the mean defect spacing.  By contrast, in high elasticity nematics, the bands can be much longer than the mean defect spacing and become more prominent features. While variable elasticity clearly changes the structures present we also quantified the differences in these active nematics by measuring their elastic energy distributions, $E_{splay} = \lvert\big(\bm{\nabla} \cdot \textbf{n})\rvert^2da$ and $E_{bend} = \lvert\big(\bm{\nabla} \times \textbf{n})\rvert^2da$, for splay and bend respectively. Fig \ref{fig1}E shows the probability distribution function (PDF) of local splay and bend distortions in the nematic. For the low elasticity nematic ($K_{11}$ = 0.26 pN, $K_{33}$ = 0.13 pN), the PDFs of the splay and bend energies both decay as power law with an exponent $n\simeq 1.7$, consistent with hydrodynamic simulations for active nematics. For the higher elasticity nematic ($K_{11}$ = 0.52 pN, $K_{33}$ = 1.04 pN), the PDFs for both splay and bend distortions deviate significantly from the power law at high energies indicating that this high elasticity nematic is more likely to store energy structurally than its low elasticity cousin.

\begin{figure*}[h]
	%\hspace{+0.3cm}
	\centerline{\includegraphics[width=0.8\textwidth]{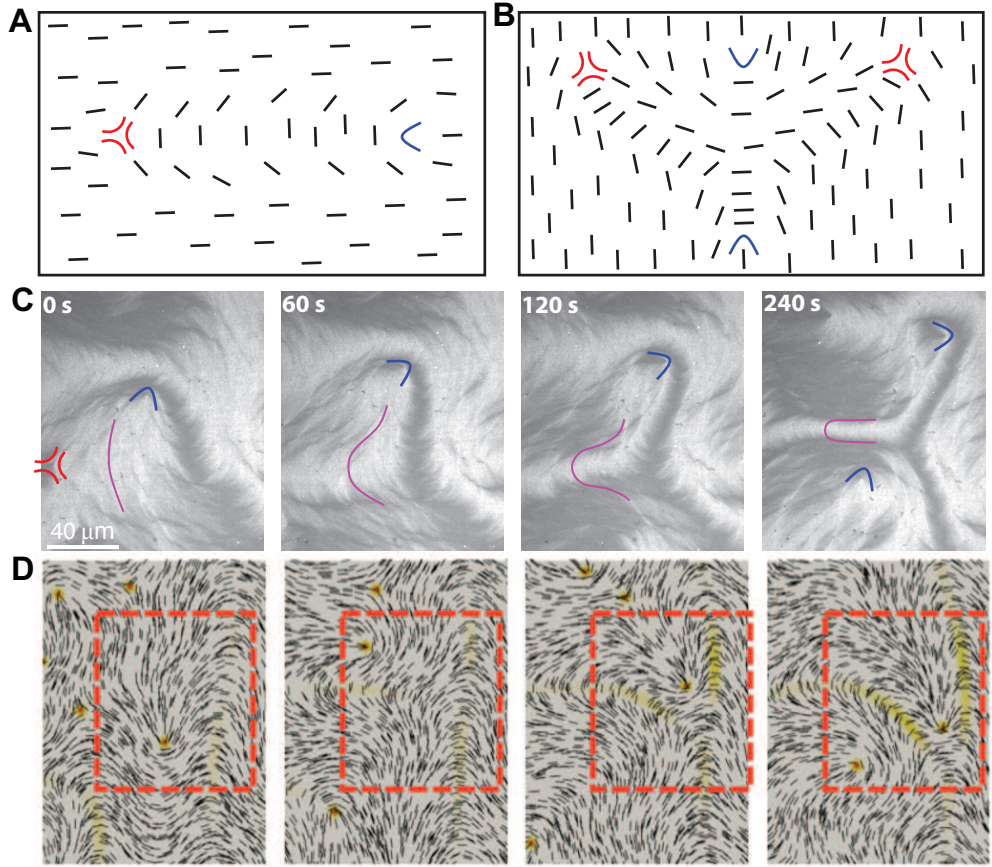}}
	\caption{\textbf{Formation and structure of branched bands}: (\textbf{A}) Schematic of topologically neutral structure containing a simple band. (\textbf{B}) Schematic of a topologically neutral structure of a branched band. Note how the +1/2 defect that is not attached to the band opposes the branch point to maintain topological neutrality. (\textbf{C}) Time series of experimental snapshots of a branched band forming in a nematic with ($K_{11}$ = 0.52 pN, $K_{33}$ = 1.04 pN). An initially aligned region (purple line) adjacent to an extant band buckles towards a -1/2 defect (red trefoil). As the bend distortion increases, a +1/2 defect appears opposing the branch point similar to the schematic in (B). (\textbf{D}) Time series snapshots of a band forming in hydrodynamic simulation. As the local bend distortion (yellow color) in the nascent band increases, a +1/2 defect approaches the branch point, stabilizing the structure. }
	\label{fig2}
\end{figure*}

In addition to the increased band length, we also observe an increase in their structural complexity. Previous observations of similar structures have noted not only the simple structures such as the schematic shown in Fig. \ref{fig2}A but also branching phenomena such as that shown in Fig. \ref{fig2}B \cite{sokolov2019}. These schematics of topologically neutral structures indicates how a stable branch might be formed. In its simplest form a band is anchored on either side by a $\pm$ 1/2 defect pair. Thus even though the length of the band is perpendicular to the surrounding director field, the structure as a whole is topologically neutral. The schematic in Fig. \ref{fig2}B demonstrates that if a band were to branch, the three free ends would all be anchored by defects as one might expect, but to maintain local topological neutrality a fourth defect that is not "attached" to the band must nonetheless be associated with the structure and oppose the branch point. 

Such a prediction bears out when we observe the branching of a band in experiment. Fig. \ref{fig2}C shows a series of experimental snapshots where an initial band, anchored on one side by a +1/2 defect, branches towards the -1/2 defect initially visible on the left hand side of the image. This branching proceeds from an initially aligned region (purple line) which appears to buckle as the movie proceeds, steadily reducing in width. A similar pattern of region buckling and band width reduction can be seen in hydrodynamic simulations shown in Fig. \ref{fig2}D where the color underneath the director field is a measure of nematic order with a darker color indicating less ordered regions. In this series of simulation snapshots we can see, just as in the experiment, an initially aligned region begin to buckle and the resultant band reduce in width over time. In both experiment and simulation, the final mature branched structures exhibit +1/2 defects pointed "towards" the branch point as in the schematic in Fig. \ref{fig2}B. Having examined the geometry of a band we now ask how the mechanics of the LC might contribute to their formation.

%\subsection{An increase in bend modulus alone suffices the band formation and banding phase diagram}
\begin{figure*}
	%\hspace{+0.3cm}
	\centerline{\includegraphics[width=0.8\textwidth]{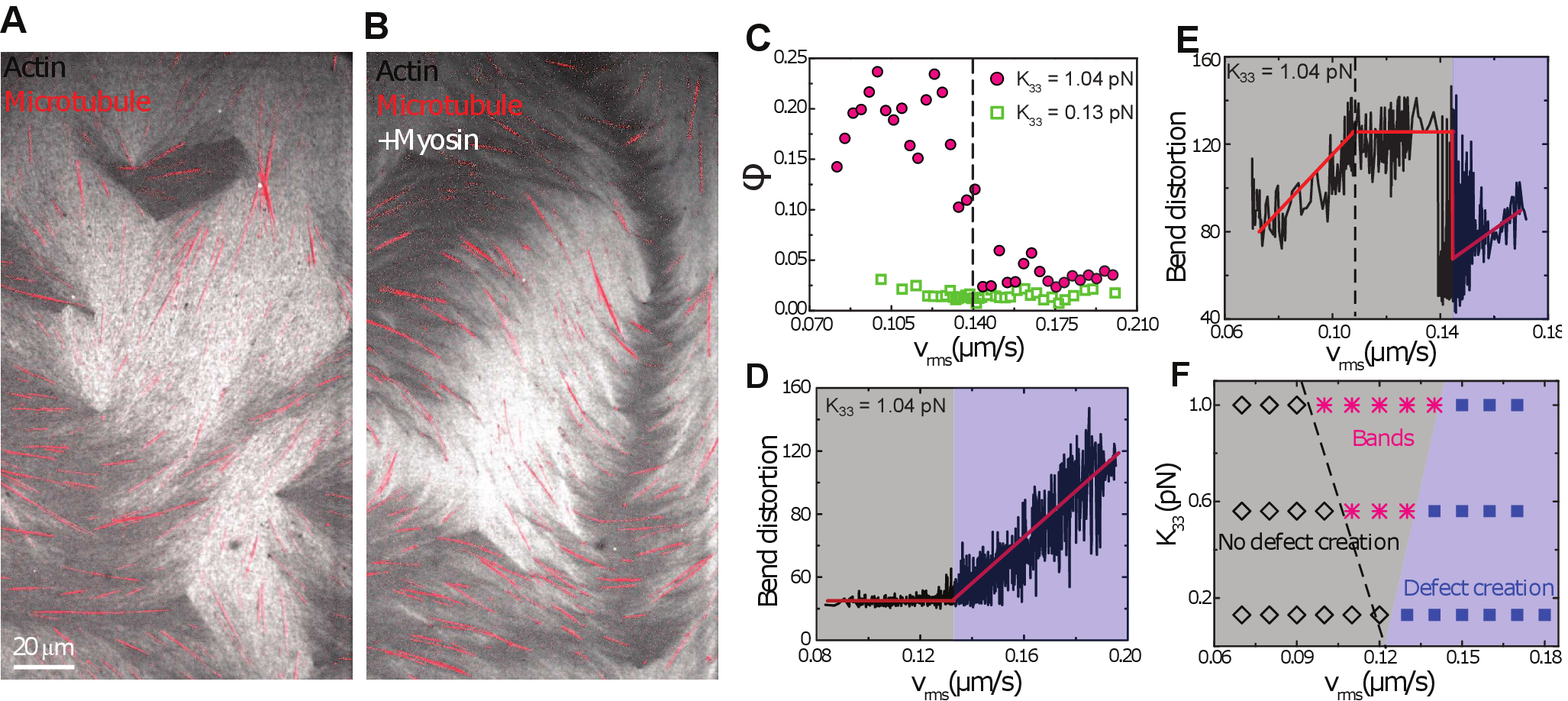}}
	\caption{\textbf{Increased bend elasticity promotes band formation at intermediate activities}: (\textbf{A}) Optical images of actin LC containing a sparse concentration of microtubules in red with bend modulus $K_{33}$ = 1.04 pN. (\textbf{B}) The same nematic as in (A) after the addition of myosin-II. Note the long band reminiscent of Fig. \ref{fig1}C. (\textbf{C}) Area fraction of bands ($\phi$) plotted as a function of root mean squared velocity, ($v_{rms}$). While for low $K_{33}$ no significant population of bands is observed, nematics with high elasticity only exhibit large area fractions of bands below a certain critical $v_{rms}$ (black dashed line). (\textbf{D}) Bend distortion ($E_{bend} = \lvert\big(\bm{\nabla} \times \textbf{n})\rvert^2da$) in a high elasticity nematic plotted as a function of $v_{rms}$. The purple region corresponds to the right hand side of the black line in (C). The black dased line here corresponds to the onset of an energetic plateau. (\textbf{E}) Total bend elastic distortion plotted as a function of $v_{rms}$ for $K_{33}$ = 0.13 pN. The purple region beyond 0.12 $\mu$m/s indicates the region of defect creation. (\textbf{F}) State diagram summarizing the dynamic states observed as a function of $K_{33}$ and $v_{rms}$. The data points are from three samples $K_{11}$=0.26 and varying $K_{33}$ over a range of activities and are color coded for regimes with no defect creation (open black diamonds), elastic bands (magenta stars) and defect creation (blue squares).  The transition from gray to blue shading indicates a crossover to active stresses sufficient for defect creation. }
	\label{fig3}
\end{figure*}

To isolate the role of bending modulus in the formation of elasticity bands, we build a composite LC, formed by a low elasticity actin nematic ($l$=1 $\mu$m) sparsely doped with microtubules (1:84 molar ratio) and shown in Fig. \ref{fig3}A. As previously described, the inclusion of microtubules, due to their higher bending rigidity, increases $K_{33}$  without influencing $K_{11}$ \cite{RZhang}. Thus, while the undoped nematic had splay and bend moduli of $K_{11}$=0.26 pN and $K_{33}$=0.13 pN respectively, the doped sample exhibited  $K_{11}$ = 0.26 pN and $K_{33}$ = 1.04 pN. This difference is particularly striking when activity is added. Whereas the undoped sample featured bands only of approximately the length of average defect spacing (Fig. \ref{fig1}B), the doped sample sample supported bands that were much longer (Fig. \ref{fig3}B, Movie S3). Active nematic dynamics arise from the competition between elasticity and activity. Thus, in order to fully understand the consequences of increased bend rigidity we consider the state of the nematic over a range of activities.

 Thus far we have focused only on a narrow range of activities. We now exploit the time-dependent variation in motor concentration to explore how changes in activity impact band formation \cite{NKumar}. We track the magnitude of active flows measured in terms of root mean squared velocity, $v_{rms}$ = $\sqrt {\langle v^2 \rangle}$ where $v$ is the local velocity measured by particle imaging velocimetry (PIV) as a function of time. Since $v_{rms}$ is proportional to active stress, the activity in our system can be conveniently expressed in terms of $v_{rms}$ \cite{YeomansNatCom,NKumar}. To quantify the magnitude of banding in our experiments, we define an order parameter $\phi$ as the area occupied by bands in the experimental field of view divided by the total area (see Methods). This order parameter is low for all velocities in the nematic with low bend elasticity (Fig. \ref{fig3}C). By contrast, when $K_{33}$ = 1.04 pN this order parameter is high for low velocities and then drops suddenly to a lower value for $v_{rms}$ $>$ 0.14 $\mu$m s$^{-1}$.

To further understand this, we track the amount of bend distortion as a function of $v_{rms}$. For the low $K_{33}$, this distortion is minimal for the lowest velocities and, beyond a threshold $v_{rms}$ shows a monotonic increase (Fig. \ref{fig3}D). The threshold beyond which distortion increases is the minimal energy needed for new defect creation.  For the high $K_{33}$ case, we find that elastic distortion increases even at low $v_{rms}$ and, there exists a range of activities where the elastic distortion plateaus (Fig. \ref{fig3}E). This plateau corresponds with the range of activities at which bands are a long lived phenomenon. At even higher activities, the bend distortion abruptly drops. Here the activity is high enough that defect nucleation dominates the dynamics and any bands that form are short lived. This suggests that, for nematics with high bend elasticity, bands are an energy barrier which the nematic must cross before reaching the well known turbulent regime.  This is in contrast to nematics with low bend elasticity where there is a direct transition to turbulence without ever creating long bend deformations. We summarize these results in Fig. \ref{fig3}F through a state diagram of $K_{33}$ and $v_{rms}$ which shows that the width of the stable band regime is strongly dependent on the strength of bend elasticity $K_{33}$. This phase diagram--while not precisely the same--bears notable resemblance to one previously predicted from simulation \cite{srivastava2016negative}. To further explore the stability of bands we turn to hydrodynamic simulations of active nematics.

\begin{figure*}[h]
	%\hspace{+0.3cm}	
	\centerline{\includegraphics[width=0.8\textwidth]{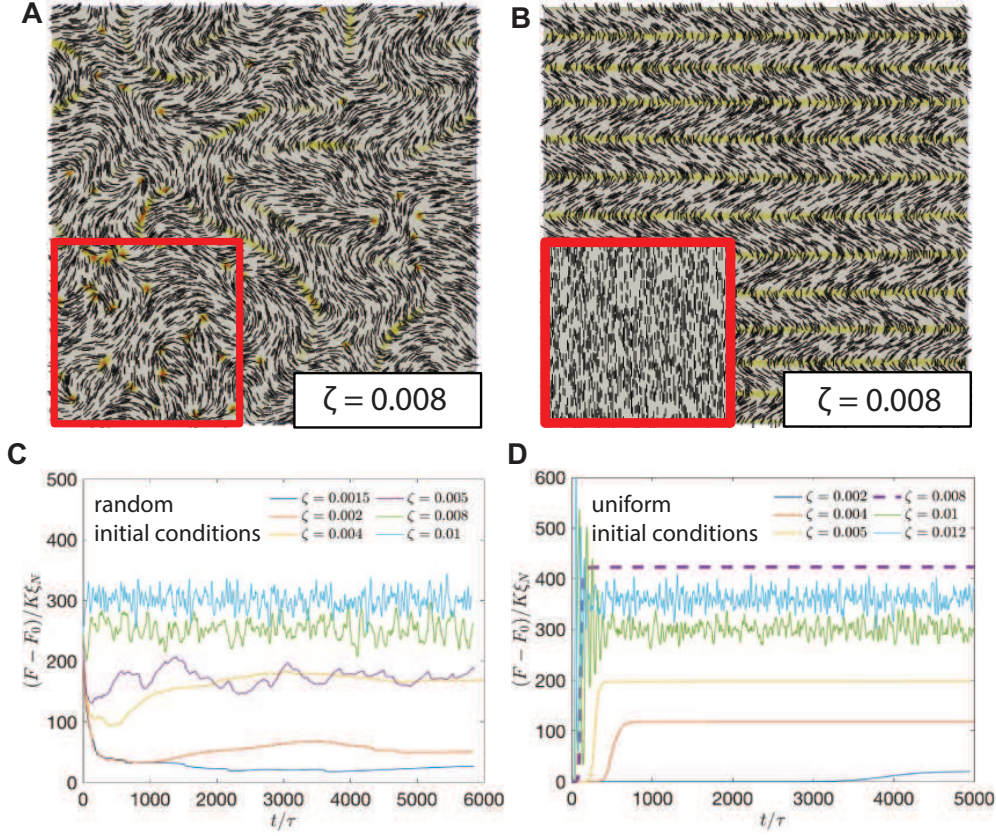}}
	\caption{\textbf{Band stability is a history dependent phenomenon}: (\textbf{A}) Simulation snapshot of fully developed active turbulence stemming from random initial conditions (inset). With these initial conditions, bands (yellow) are relatively short. (\textbf{B}) Simulation snapshot of stable structures formed in a nematic at the same activity level stemming from uniform initial conditions (inset). Note that the bands arising from these initial conditions are a stable series of oppositely directed bands that span the space of the simulation box. (\textbf{C}) Steady state elastic energy plotted over time for simulations with random initial conditions across a range of activities. (\textbf{D}) Steady state elastic energy for simulations stemming from uniform initial conditions. The dashed purple curve is the highest activity level that does not produce defect pairs and only exhibits stable bands. }
	\label{fig4}
\end{figure*}

The model we employ here has proven successful in understanding many of active nematic phenomena in lyotropic systems, including actin, microtubule and living-liquid-crystal systems\cite{zhang2016,NKumar,sokolov2019}. To understand the stability of elasticity bands and their relationship with defects, we prepare the initial director field in two ways, one random and the other uniform (Fig.~\ref{fig4}A,B, inset). For a random initial configuration, simulations show that at sufficient high activity when active turbulence is fully developed, short and transient elasticity bands are seen (Fig.~\ref{fig4}A, Movie S4). In contrast, at the same activity level, a simulation instantiated with uniform initial conditions exhibits a series of oppositely directed bands that span the simulation box (Fig.~\ref{fig4}A). To understand the differences between these two cases we plot the system's steady-state elastic energy as a function of time (Fig.~\ref{fig4}C,D). We find that in simulations stemming from random initial conditions two basic trends hold. First, we observe that the average steady state value of the elastic energy monotonically increases with activity. Similarly, we find that the frequency of fluctuations in elastic energy increases with activity (Fig.~\ref{fig4}C). These trends can be understood simply in terms of defect dynamics. As activity increases, the average number of defects at steady state increases while the frequency of creation and annihilation events experiences a similar increase. With this in mind, it is notable that in simulations stemming from uniform initial conditions, the average steady state energy is in fact not a monotonic function of activity (Fig.~\ref{fig4}D). In fact, what we see is that the elastic energy increases with activity to a point (dashed black line) and then decreases. This decrease corresponds to the emergence of fluctuations in the steady state elastic energy and thus the emergence of defects. Previous studies have pointed out that elasticity bands are formed from a uniform director field due to bend-driven hydrodynamic instabilities, and are precursors to the defect state. Interestingly, our results show that the stability of bands in simulations with uniform initial conditions is a consequence of the symmetry of the initial state and that breaking this symmetry leads to the breakdown of bands at much lower activities. This breakdown of bands is due to their decomposition into and interactions with topological defects which have very interesting consequences for the dynamics of the system. 

\begin{figure*}[h]
	%\hspace{+0.3cm}	
	\centerline{\includegraphics[width=0.8\textwidth]{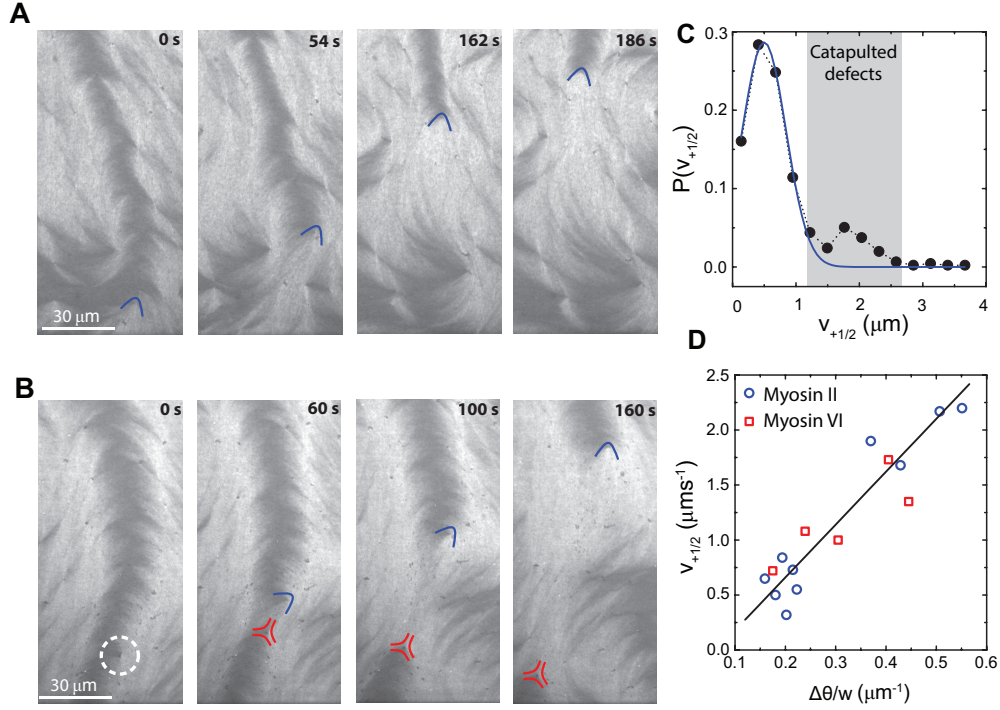}}
	\caption{\textbf{`Catapulting' of +1/2 defects through elasticity bands}: (\textbf{A})Time-series of fluorescent images of actin LC ($K_{11}$ = 0.52 pN, $K_{33}$ = 1.04 pN) showing a +1/2 defect (blue chevron) moving along a band leaving a uniformly aligned region in its wake. (\textbf{B}) Time series of images showing a band severing event. The band separates near its thinnest point into a $ \pm$ 1/2 defect pair after which dynamics proceed as in (A). (\textbf{C}) Distribution of speeds of +1/2 defects in the movies from which A $\&$ B originated. The blue line coreresponds to a Gaussian fit of the data (black circles). The second peak (outlined in gray) deviates significantly from this single Gaussian fit and corresponds to defects we term 'catapulted'. (\textbf{D}) The speed of +1/2 defects within bands tracked over multiple experiments. Defect speed scales inversely with the band strength defined in Fig. \ref{fig1}F. The data includes actin LCs driven by both myosin-II (red squares) as well as synthetic myosin-VI motors (blue circles). The solid line is the fit to the analytical model.}
	\label{fig5}
\end{figure*}

Bands are highly dynamic and, much like defects, undergo spontaneous creation and annihilation. While bands are created from a 'buckling' of the director field as in Fig. \ref{fig2}C, they are annihilated by the motion of defects. Due to their compatible geometries, a +1/2 defect moving along a band--such as the schematic Fig. \ref{fig2}A--will lead to a shortening of the band. This can be clearly seen in the experimental snapshots in Fig. \ref{fig5}A where a +1/2 defect moves along a band leaving an aligned director field in its wake (Movie S5). Band shortening in this manner can also proceed after a severing event. As bands mature, they thin to a width set by the competition between activity and elasticity. If bands become sufficiently thin, they can sever via the formation of a $\pm$ 1/2 defect pair as seen in Fig. \ref{fig5}B (Movie S6). These nascent defect pairs move quickly through the band leading to rapid annihilation after nucleation. To understand this phenomenon, we plot the probability density of +1/2 defect speeds in the nematic. We find that while most defects are distributed gaussianly around a single speed, a small population move significantly faster resulting in a bimodal distribution of defect speeds, Fig. \ref{fig5}C (Movie S7). This anomalous defect speed is the result of the release of stored elastic energy from the band in the form of defect motion. To see this point concretely we plot the speed of defects moving through bands as a function of band strength, $\Delta\theta/w$, as defined in Fig. \ref{fig1}F, in Fig. \ref{fig5}D. The data from experiments with myosin-II can be seen in blue and clearly show that thinner bands result in faster defect propulsion. To ensure that such an effect is indeed due to the bands themselves and is not in fact a product of motor driven activity, we perform experiments with a synthetic myosin-VI motor \cite{schindler2014engineering}. We find that band-associated defects in this system follow the same trend as in the myosin-II nematic (Movie S8). This indicates that the catapulting of defects through bands is indeed a product of stored bend elasticity in the material and not the specifics of motor propulsion. In this manner, the presence of bands in the nematic system leads to a novel dynamical steady state where defects are propelled not only by molecular motors but also by stored elastic energy.

\section{Conclusions}

Here we explored the effect of nematic bend elasticity on emergent  flows in active nematics. By utilizing experimental control over specifically bend elasticity, we found a novel dynamical phase where defects and elongated elasticity bands coexist and interact to alter the nature of complex flows. We described the structure of these bands in terms of their width and the change in director angle from one side to the other. Furthermore, we saw that bands can exist not only in their simplest linear form but also in higher order branched structures. We showed that increases in the bend elastic modulus are sufficient to control band formation. This further strengthens previous findings that disparate elastic constants (i.e. $K_{11} \neq K_{33}$) not only play an important role in determining the shape of topological defects \cite{RZhang} and defect density \cite{NKumar}, but also in predicting the dynamics of complex active flows. 

Previous work has suggested that active nematics are endowed with a fundamental length scale which is given by $l_c$ = $\sqrt{K/\alpha}$ where $K$ and $\alpha$ are magnitudes of elasticity and local active stress, respectively. In this framework, an increased $K$ corresponds to a higher energy barrier required to undergo defect formation. While our system does not map perfectly onto this description due to disparate elastic constants, we nonetheless see similar phenomena with bands. Bands are less prevalent at high activities and low bend elasticities. Furthermore, at intermediate activity and high bend elasticity, a state occurs in which bands are long lived, corresponding to the increased barrier for defect creation at these high elasticity values. When the activity is increased further however we find a state that is dominated by defects, akin to the active turbulence described previously. 

Bands are not a stranger in the active nematic literature, however discrepancies have existed between the long lived structures described in simulations at low activity (Fig.\ref{fig4}D) and the transient structures in experiments (Fig.\ref{fig5}B). Here we showed that this discrepancy can be explained by initial conditions. We found that simulations started from uniform initial conditions supported indefinitely stable bands over a very large range of activities and resulted in a decrease in stored elastic energy upon their disintegration. Meanwhile, simulations that proceeded from random initial conditions showed monotonic energy scaling and shorter lived bands. This further underlines how bands are not merely a structural quirk but an important player in the LC dynamics. 

When bands are present in a system with intermediate activity, they play a central role in the dynamics. For one, bands dictate the path of +1/2 defects in the system as they 'zips' up bands, leaving a uniformly aligned director field in their wake. Furthermore, a band itself can not only nucleate a defect pair but also convert stored bend elastic energy into motion in the form of catapulting defects. The combination of these effects leads to an exotic dynamical state when bend elasticity is about equal in strength to extensile activity. 

Overall then, bands are an example of a situation in which the interplay of nematic elasticity and activity produces a novel dynamical state. While the state described in this work is due to the competition between specifically bend elasticity and extensile stress, one could imagine that different combinations of stress modalities and specific elasticities might result in many interesting states.

%Another important finding is that, for an extensile active nematic, the only elastic mode which plays significant role in resisting or competing active stress is the bend modulus ($K_{33}$).  This is evident from elasticity bands playing an important role in dictating the speed and direction of +1/2 topological defects, the effect named here as catapulting. This result also shows how deterministic characters evolve of otherwise complex turbulent flows at low activity levels which might indicate a signature of a state before the fully turbulent state is achieved. Further experiments and analysis are required to quantify this transition to turbulence and we defer this to future work.

\section*{Author Contributions}
N.K., R.Z., J.dP., and M.G. conceptualized the work. N.K. performed the experiments. R.Z. performed the simulations. S.R. collected additional data. All authors wrote and edited the manuscript. 

\section*{Conflicts of interest}
There are no conflicts to declare.

\section{Acknowledgments}
N.K. and R.Z. thank Sumesh Thampi, Sriram Ramaswamy, Christina Marchetti, Daniel Needleman and Zvonimoir Dogic for useful discussions. NK thanks Dr. Kimberly Weirich and Dr. Samantha Stam for purified proteins. This work was primarily supported by the University of Chicago Materials Research Science and Engineering Center, which is funded by National Science Foundation under award number DMR-2011854. M.L.G.  acknowledge support from NSF Grant DMR-1905675. J.J.d.P. acknowledges support from NSF Grant DMR-1710318. N.K. acknowledges the Yen
Fellowship of the Institute for Biophysical Dynamics, The University of
Chicago. R.Z. is grateful for the support of the University of Chicago Research Computing Center for assistance with the calculations carried out in this work

%%%END OF MAIN TEXT%%%

%%%REFERENCES%%%
\bibliography{MSBands.bib}

\section*{Methods}
\subsection{Experimental methods}
\subsubsection{Proteins}
We purify monomeric actin purified from rabbit skeletal muscle acetone powder purchased from Pel-Freez Biologicals, Rogers, AR) \cite{actin} and stored at $-80^0$C in G-buffer containing 2mM Tris HCL pH 8.0, 0.2 mM APT, 0.2 mM CaCl$_{2}$, 0.2 mM DTT, $0.005\%$ NaN$_{3}$). For fluorescence microscopy, we label G-actin with Tetramethylrhodamine-6-maleimide dye (Life Technologies, Carlsbad, CA). Capping protein (with a His-tag) is used to regulate actin filament length purified from bacteria (plasmid gifted by Dave Kovar lab, The University of Chicago) \cite{CP}. Mictotubules are polymerized in PEM-100 buffer at 37$^\circ$ C (100mM Na-PIPES, 1mM MgSO$_{4}$, 1mM EGTA, pH 6.8) in the 1:10 ratio of fluorescently labeled tubulin (Cytoskeleton, cat$\#$ TL488M) and unlabeled tubulin (Cytoskeleton, cat$\#$ HTS03) in the presence of 1 mM GMPCPP (Jena Biosciences, cat $\#$ NU-405L). They are later stabalized by adding 50 $\mu$M Taxol. The microtubule length is shortened by shearing through Hamilton Syringe (Mfr $\#$ 81030, Item $\#$ EW-07939-13). Skeletal muscle myosin II is purified from chicken breast \cite{Myo1} and labeled with Alexa-642 maleimide (Life Technologies, Carlsbad, CA) \cite{Myo2}. Synthetic myosin-VI motors were purified from the construct $M6DI_{816}2R_TET$ from \cite{schindler2014engineering} and were a gift from the lab of Zev Bryant. 

\subsubsection{Experimental assay and microscopy}
Actin from frozen stocks stored in G-buffer is added to a final concentration of 2 $\mu$M with a ratio 1:5 TMR-maleimide labeled:unlabeled actin monomer. We polymerize actin in 1X F-buffer (10 mM imidazole, pH 7.5, 50mM KCL, 0.2mM EGTA, 1mM MgCl$_{2}$ and 1mM ATP). To minimize photobleaching, an oxygen scavenging system (4.5 mg/mL glucose, 2.7 mg/mL glucose oxidase(cat$\#$345486, Calbiochem, Billerica, MA), 17000 units/mL catalase (cat $\#$02071, Sigma, St. Louis, MO) and 0.5 vol. $\%$ $\beta$-mercaptaethanol is added. We use 0.3 wt $\%$ 15 cP methylcellulose as the crowding agent for actin filaments. Frozen capping protein stocks are thawed on ice and are added at the same time (6.7 and 3.3 nM for 1 $\mu$m and 2 $\mu$m long actin filaments respectively). Myosin-II is mixed with phalloidin-stabilized F-actin at a 1:4 myosin/actin molar ratio in spin-down buffer and centrifuged for 30 min at 100,000 × g. The supernatant containing myosin with low affinity to F-actin is used in experiments whereas the high-affinity myosin is discarded. For experiments with microtubules, taxol-stabilized microtubules are added to the final concentration of 1 $\mu$g/mL, a 1:84 ratio of microtubules to actin.

We use a glass cylinder (cat$\#$ 09-552-22, Corning Inc.) glued to a coverslip as an experimental sample same as \cite{NKumar}. Coverslips are cleaned by sonicating in water and ethanol. To create a hydrophobic surface, they are further treated with triethoxy(octyl)silane in isopropanol. For creating an oil-water interface, PFPE-PEG-PFPE surfactant (cat $\#$ 008, RAN biotechnologies, Beverly, MA) is dissolved in Novec-7500 Engineered Fluid (3M, St Paul, MN) to a concentration of 2$\%$ wt/volume. To prevent bulk flows at the surface, a small $2\times2$ mm teflon mask is placed on the treated coverslip before exposing it to UV/ozone for 10 minutes. A glass cylinder thouroughly cleaned with water and ethanol and is glued to the coverslip using instant epoxy. $3~\mu$L of oil-surfactant solution is added into the chamber, and quickly pipetted out (3 s) to leave a thin oil coating. The polymerization mixture is immediately added afterwards. 30$-$60 minutes later, a thin layer of actin LC is formed. The sample is always imaged in the middle of the film over the camera field of view, which is approximately 200 $\mu$m $\times$ 250 $\mu$m to make sure that the sample remains in focus over this area, which is far away from the edges.  Myosin II \textbf{or tetrameric myosin} motors are added to the polymerization mixture at 5$-$10 nM and 200 pM respectively.

The sample is imaged using an inverted microscope (Eclipse Ti-E; Nikon, Melville, NY) with a spinning disk confocal head (CSU-X, Yokagawa Electric, Musashino, Tokyo, Japan), equipped with a CMOS camera (Zyla-4.2 USB 3; Andor, Belfast, UK). A 40X 1.15 NA water-immersion objective (Apo LWD; Nikon) was used for imaging. Images were collected using 491 nm, 568 nm and 642 nm excitation for microtubules, actin, and myosin-II respectively. Image acquisition was controlled by Metamorph (Molecular Devices, Sunnyvale, CA).

\subsubsection{PIV and root-mean-squared velocity}
The active flows are quantified using particle image velocimetry (available at \url{www.oceanwave.jp/softwares/mpiv/}) to extract local velocity field, $\textbf{v}$. The images were processed through unsharped masking and then background subtraction using built-in plugins in ImageJ software \cite{imagej}. The grid size of 2.4 $\mu$m was used for PIV vector calculation and images were separated by a time-interval of 5 s. 

\subsubsection{Band Characterization and banding order parameter}
To identify bands in fluorescent images, we perform a series of image processing algorithms in ImageJ software. We first enhance the contrast of our images using the CLAHE plugin. Later we use "shape index map" plugin to separate the bands which are later outlined using "edge detection" plugin. This image is then thresholded which marks the band outlines and separate them from the rest of the image. Using this, we then calculate the area inside the bands and divide it by the total area to calculate $\phi$.

\subsection{Numerical methods}
\subsubsection{Continuum Model}
The total free energy of the nematic LC, $F$, consists of a bulk and a surface term:
\begin{equation}\label{total}
\begin{aligned}
F&= \int_{V} dV f_{bulk} + \int_{\partial V} dS f_{surf} \\
&= \int_{V} dV (f_{LdG}+f_{el}) + \int_{\partial V} dS f_{surf},
\end{aligned}
\end{equation}
where $f_{LdG}$ is the short-range free energy, $f_{el}$ is the long-range elastic energy, and $f_{surf}$ is the surface free energy associated with preferred nematic orientation. $f_{LdG}$ is the Landau-de Gennes in the Doi form\cite{deGennes_book}:
\begin{equation}\label{phase}
f_{LdG}= \frac{A_0}{2} (1-\frac{U}{3} )\tr({\bf Q}^2) - \frac{A_0U}{3} \tr({\bf Q}^3) + \frac{A_0 U}{4}(\tr({\bf Q}^2))^2.
\end{equation}
Parameter $U$ controls the magnitude of $q_0$, namely the equilibrium scalar order parameter via $q_0=\frac{1}{4}+\frac{3}{4}\sqrt{1-\frac{8}{3U}}$.
The elastic energy $f_{el}$ is written as ($Q_{ij,k}$ means $\partial_k Q_{ij}$):
\begin{equation}\label{elastic_en}
\begin{aligned}
f_{el}=&\frac{1}{2}L_1 Q_{ij,k}Q_{ij,k}+\frac{1}{2}L_2 Q_{jk,k}Q_{jl,l}\\
&+\frac{1}{2}L_3 Q_{ij}Q_{kl,i}Q_{kl,j}+\frac{1}{2}L_4 Q_{ik,l}Q_{jl,k}.
\end{aligned}
\end{equation}
If the system is uniaxial, the above equation is equivalent to the Frank Oseen elastic energy expression:
\begin{equation}\label{frank}
\begin{aligned}
f_e= \frac{1}{2}K_{11} &(\nabla\cdot {\bf n})^2+\frac{1}{2}K_{22}({\bf n}\cdot \nabla \times{\bf n})^2+\frac{1}{2}K_{33}({\bf n}\times (\nabla \times{\bf n}))^2 \\
& -\frac{1}{2}K_{24}\nabla\cdot [{\bf n}(\nabla\cdot{\bf n})+{\bf n}\times(\nabla\times{\bf n})].
\end{aligned}
\end{equation}
The $L$'s in Eq.~\ref{elastic_en} can then be mapped to the $K$'s in Eq.~\ref{frank} via
\begin{equation}
\begin{aligned}
L_1&=\frac{1}{2q_0^2} \left[ K_{22}+\frac{1}{3}(K_{33}-K_{11})  \right], \\
L_2&=\frac{1}{q_0^2} (K_{11}-K_{24}), \\
L_3&=\frac{1}{2q_0^3} (K_{33}-K_{11}), \\
L_4&=\frac{1}{q_0^2} (K_{24}-K_{22}).
\end{aligned}
\end{equation}
Point wise, ${\bf n}$ is the eigenvector associated with the greatest eigenvalue of the ${\bf Q}$-tensor at each lattice point.

To simulate active LC's dynamics, a hybrid lattice Boltzmann method is used to simultaneously solve a Beris-Edwards equation and a momentum equation which accounts for the hydrodynamic flows.
By introducing a velocity gradient $W_{ij}=\partial_j u_i$, strain rate $\bf A=(\bf W + \bf W^T)/2$, vorticity $\bf \Omega =(\bf W - \bf W^T)/2$, and a generalized advection term
\begin{equation}
\begin{aligned}
{\bf S}({\bf W},{\bf Q})=&(\xi {\bf A}+{\bf \Omega})({\bf Q}+{\bf I}/3)+({\bf Q}+{\bf I}/3)(\xi {\bf A}-{\bf \Omega})&\\
&-2\xi ({\bf Q}+{\bf I}/3) \tr({\bf Q W}),&
\end{aligned}
\end{equation}
one can write the Beris-Edwards equation\cite{beris_edwards_book} according to
\begin{equation} \label{beris_edwards_eq}
(\partial_t +{\bf u}\cdot \nabla){\bf Q}-{\bf S}({\bf W},{\bf Q})=\Gamma \bf{H}.
\end{equation}
The constant $\xi$ is related to the material's aspect ratio, and $\Gamma$ is related to the rotational viscosity $\gamma_1$ of the system by $\Gamma=2q_0^2/\gamma_1$\cite{denniston_2d}. The molecular field $\bf{H}$, which drives the system towards thermodynamic equilibrium, is given by
\begin{equation}
{\bf H}=-\left[ \frac{\delta F}{ \delta \bf{Q}} \right]^{st},
\end{equation}
where $\left[ ...\right]^{st}$ is a symmetric and traceless operator. When velocity is absent, i.e. ${\bf u}({\bf r})\equiv 0$, Besris-Edwards equation Eq.~\ref{beris_edwards_eq} reduce to Ginzburg-Landau equation:
$$
\partial_t {\bf Q}=\Gamma {\bf H}.
$$
To calculate the static structures of $\pm 1/2$ defects, we adopt the above equation to solve for the ${\bf Q}$-tensor at equilibrium.

Degenerate planar anchoring is implemented through a Fournier-Galatola expression\cite{fournier_galatola} that penalizes out-of-plane distortions of the $\bf{Q}$ tensor. The associated free energy expression is given by
\begin{equation}
f_{surf} = W (\tilde{\bf Q} - \tilde{\bf Q}^{\perp})^2,
\end{equation}
where $\tilde{\bf{Q}} = \bf{Q} + (q_0/3)\bf{I}$ and $\tilde{\bf{Q}}^{\perp} = {\bf{P}} \tilde{\bf{Q}} {\bf{P}}$. Here ${\bf P}$ is the projection operator associated with the surface normal ${\bf \nu}$ as ${\bf P}=\bf{I} -{\bf \nu} {\bf \nu} $. The evolution of the surface ${\bf Q}$-field is governed by\cite{zhang_jcp2016}:
\begin{equation} \label{surface_evolution}
\frac{\partial {\bf Q}}{\partial t}=-\Gamma_s \left( -L {\bf \nu} \cdot \nabla {\bf Q} +\left[  \frac{\partial f_{surf}}{\partial {\bf Q}} \right]^{st} \right),
\end{equation}
where $\Gamma_s=\Gamma/\xi_N$ with $\xi_N=\sqrt{L_1/A_0}$, namely nematic coherence length.

Using an Einstein summation rule, the momentum equation for the nematics can be written as\cite{denniston_2d,denniston_3d}
\begin{equation}  \label{ns_eq}
\begin{aligned}
\rho(\partial_t+u_j \partial_j)u_i=&\partial_j \Pi_{i_j}+\eta\partial_j[\partial_i u_j+\partial_j u_i +(1-3\partial_\rho P_0)\partial_\gamma u_\gamma \delta_{i_j}].
\end{aligned}
\end{equation}
The stress $${\bf{\Pi}}={\bf{\Pi}}^{p}+{\bf{\Pi}}^{a}$$ consists of a passive and an active part. The passive stress ${\bf \Pi}^{p}$ is defined as
\begin{equation}
\begin{aligned}\label{stress}
\Pi_{i j}^{p}=  & -P_0 \delta_{i j}-\xi H_{i \gamma} (Q_{\gamma j} +\frac{1}{3}\delta_{\gamma j}) - \xi (Q_{i \gamma} +\frac{1}{3} \delta_{\gamma j} ) H_{\gamma j} & \\
 &+ 2 \xi (Q_{i j} +\frac{1}{3}\delta_{i j}) Q_{\gamma \epsilon}H_{\gamma \epsilon} -\partial_{j} Q_{\gamma \epsilon} \frac{\delta \mathcal F}{\delta \partial_i Q_{\gamma \epsilon}} &\\
 & + Q_{i \gamma} H_{\gamma j} -H_{i \gamma} Q_{\gamma j}, &
\end{aligned}
\end{equation}
where $\eta$ is the isotropic viscosity, and the hydrostatic pressure $P_0$ is given by\cite{fukuda_force}
\begin{equation}
P_0=\rho T - f _{bulk}.
\end{equation}
The temperature $T$ is related to the speed of sound $c_s$ by $T=c_s^2$. The active stress reads\cite{yeomans_pre2007}
\begin{equation} \label{active_stress}
\Pi_{i j}^{a}=  -\alpha Q_{i j},
\end{equation}
in which $\alpha$ is the activity in the simulation. The stress becomes extensile when $\alpha>0$ and contractile when $\alpha<0$.

\subsubsection{Numerical Details}\label{m5}
We solve the evolution equation Eq.~\ref{beris_edwards_eq} using a finite-difference method. The momentum equation Eq.~\ref{ns_eq} is solved simultaneously via a lattice Boltzmann method over a D3Q15 grid\cite{zhaoliguo_book}. The implementation of stress follows the approach proposed by Guo {\it et al.}\cite{zhaoliguo_forcing}. The units are chosen as follows: the unit length $a$ is chosen to be $a=\xi_N=1~\mu m$, characteristic of the filament length, the characteristic viscosity is set to $\gamma_1$=0.1~Pa$\cdot$s, and the force scale is made to be $F_0=10^{-11}$~N. Other parameters are chosen to be $A_0=0.1$, $K=0.1$, $\xi=0.8$, $\Gamma=0.13$, $\eta=0.33$, and $U=3.5$ leading to $q_0\approx 0.62$.
The simulation is performed in a rectangular box. The boundary conditions in the $xy$ plane are periodic with size $[N_x, N_y]=[250,~250]$. Two confining walls are introduced in the $z$ dimension, with strong degenerate planar anchoring, ensuring a quasi 2D system with z-dimension $ N_z=9$.
We refer the reader to Ref.~\cite{zhang_jcp2016} for additional details on the numerical methods employed here.

\subsection{Estimate Viscosity}\label{m6}
To understand defect's high velocity in the elastic band, we build on elasticity theory at over-dumped limit. Say an elastic band has width $b$ and depth $t$ (thickness of the 2D film). Given the one-elastic-constant $K$, the elastic force that drives the defect motion can be written as
$$
F_e=\frac{1}{2}K\pi^2t/b.
$$
The viscous drag force is written as\cite{lavrentovich_book}
$$
F_d=\pi \gamma_1 k^2 v \ln (3.6/Er) t,
$$
where $\gamma1$ is the rotational viscosity, $k$ is the topological charge of the defect, $v$ is the defect velocity, and $Er=\gamma_1 v r_c/K$ is the defect core's Ericksen number with $r_c$ the core radius. At steady state, the above two forces are equal. We find a rigorous formulus for the defect velocity:
$$
v=\frac{2 \pi K}{\gamma_1 b \ln \left( \frac{3.6}{Er} \right)}.
$$
At low Ericksen number or low defect velocity when $Er\ll 3.6$, one has $v\propto 1/b$. To extract rotational viscosity of the nematic in the experiment, we fit the measured data with the above equation with two fitting parameters $\gamma_1$ and $K$.

\end{document}